\documentclass[journal]{IEEEtran}

\IEEEoverridecommandlockouts                              

\usepackage{epstopdf}
\usepackage{graphics}
\usepackage{fancyhdr}
\usepackage{float}
\usepackage{color}

\usepackage{footnote}
\makesavenoteenv{tabular}

\usepackage{pdfpages}
\usepackage{gensymb}
\usepackage{sidecap}
\usepackage{breakurl} 
\usepackage{subfig}
\usepackage{marginnote}
\usepackage{setspace}
\usepackage{placeins}
\usepackage{listings}
\usepackage{pgfplots}
\usepackage{tabulary}

\usepackage[labelfont=bf]{caption}
\usepackage{graphicx}
\usepackage{here}
\usepackage{amsmath}
\usepackage{amssymb}
\usepackage{siunitx}
\usepackage{color} 
\usepackage{transparent} 
\usepackage[hyphens]{url}
\usepackage[breaklinks]{hyperref}
\hypersetup{
    colorlinks=true,
    linkcolor=blue,
    filecolor=magenta,      
    urlcolor=cyan,
}

\usepackage{algorithm}
\usepackage{algpseudocode}
\usepackage{mathtools}

\usepackage{tabularx}

\usepackage{cite}

\graphicspath{{./figures/}}

\begin{document}

\title{
Wisture: RNN-based Learning of Wireless Signals for Gesture Recognition in Unmodified Smartphones
}
\author{Mohamed Abudulaziz Ali Haseeb, Ramviyas Parasuraman
         \thanks{M. Abdulaziz is affiliated with the Robotics, Perception and Learning (RPL) Lab, KTH Royal Institute of Technology, Stockholm 100 44, and with Watty AB, Sweden. 
         
R. Parasuraman is with Purdue University, West Lafayette, 47906, USA.
         
email: \tt{moaah@kth.se; ramviyas@purdue.edu}}}



\maketitle
\begin{abstract}


This paper introduces Wisture, a new online machine learning solution for recognizing touch-less dynamic hand gestures on a smartphone. Wisture relies on the standard Wi-Fi Received Signal Strength (RSS) using a Long Short-Term Memory (LSTM) Recurrent Neural Network (RNN),
thresholding filters and traffic induction.
Unlike other Wi-Fi based gesture recognition methods, the proposed method does not require a modification of the smartphone hardware or the operating system, and performs the gesture recognition without interfering with the normal operation of other smartphone applications. 

We discuss the characteristics of Wisture, and conduct extensive experiments to compare its performance against  state-of-the-art machine learning solutions in terms of both accuracy and time efficiency. The experiments include a set of different scenarios in terms of both spatial setup and traffic between the smartphone and Wi-Fi access points (AP). The results show that Wisture achieves an online recognition accuracy of up to $94\%$ (average $78\%$) in detecting and classifying three hand gestures.

\end{abstract}



\IEEEpeerreviewmaketitle

\section{Introduction}

\IEEEPARstart{s}{martphones} have become the first high performance computing and sensing devices that are carried by a majority of the population in many countries.  Today, smartphones are responsible for the majority of internet search traffic~\cite{searchengineland} as well as online media consumption~\cite{techcrunch}. Consumer reports reveal that users spend more time on smartphones than they do on desktop computers~\cite{smartinsight}.  However, their inherent limitations in terms of physical size, screen size, computing capacity and battery power, calls for interaction modalities beyond the smartphone touch screens.
Therefore, researchers and engineers are continually looking for novel methods to enrich and simplify how humans interact with their smartphones, such as natural language processing \cite{Kumar_rethinkingspeech}, and touch-less gestures~\cite{sigg6813955}.

Modern gesture recognition methods mainly use two sensing modalities: intertial measurements \cite{Cohn:2012:HUB:2207676.2208330} and images from the camera \cite{Song:2014:IGA:2642918.2647373}.
The advantages of these modalities include availability and accuracy. However,
intertia-based systems require the smartphone to be held by the user, and image based techniques
 suffer from drawbacks such as limited sensing range, sensitivity to lighting conditions, and more importantly high-power consumption.
Therefore, in recent years,
Radio Frequency signals and Received Signal Strength (RSS) have been exploited  for sensing human activities and gestures \cite{Zhao:2014:SDI:2642918.2647380,Sigg:2013:LRF:2536853.2536873}, mainly due to their advantages in terms of low power consumption, and being able to handle  non line-of-sight conditions. For instance, Google's project Soli \cite{soli} aims at creating  a rich gesture recognition interface using a specialized radar device that can be embedded into wearables, phones, computers, and Internet of Things (IoT) devices. 

The ubiquitous nature of Wi-Fi technology makes it attractive  for gesture recognition in smartphones. Thus, a number of novel methods have been proposed for using Wi-Fi RSS (e.g. \cite{DBLP:journals/corr/NandakumarKG14, sigg6813955}). However, they require either special hardware or special software modifications such as root privileges, blocked data traffic to other applications, etc. to operate.

\begin{figure}[t]
	\centering
	\vspace{-3mm}
	\includegraphics[width=0.78\linewidth]{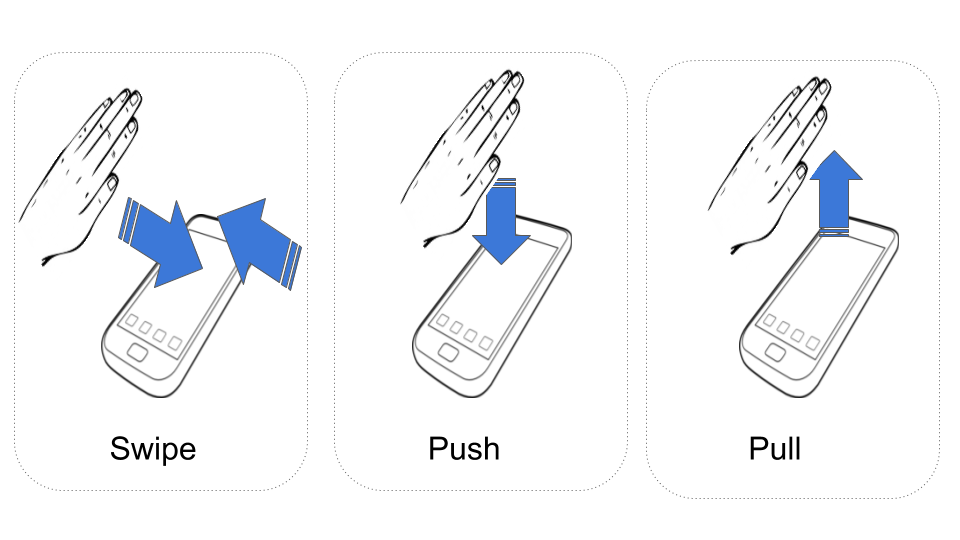}
	\vspace{-3mm}
	\caption{The hand gestures considered in this paper.}
	\label{fig:gestures}
\end{figure}
In this paper, we propose a novel machine learning method based on LSTM RNN for recognizing gestures in unmodified smartphones, based on artificially induced data traffic between the smartphone and the Wi-Fi Access Point (AP). The main contributions of this paper are:
\begin{enumerate}
\item We demonstrate that the Wi-Fi RSS can be used to recognize hand gestures near smartphones (see Fig.~\ref{fig:gestures}) using a fusion of machine learning techniques and custom signal processing algorithms.
\item We conduct several experiments under various conditions to validate the gesture classification performance of the proposed method, and compare it against state-of-the-art machine learning methods.
\item We release the experiment dataset (an extensive collection of labeled Wi-Fi RSS measurements corresponding to multiple hand gestures made near a smartphone under different spatial and data traffic scenarios.) to enable comparison with future approaches\footnote{The dataset is available at \url{https://goo.gl/2AQKdT}, and will be published to the public CRAWDAD repository (\url{http://crawdad.org}).}. 
\item We openly share the source codes of part of the Wisture\footnote{Accessible at \url{https://github.com/mohaseeb/wisture}. We also release the Android apps "Wisture" and "Winiff", available in Google Play Store.}, specifically to record high frequency Wi-Fi RSS measurements using the induced traffic approach.
\end{enumerate}

\section{Related Works}

Radio signals, notably Wi-Fi signals, have been recently exploited for sensing and recognizing human activities \cite{WANG201520,savazzi2016device}. For instance, customized hardware-based active sensing solution is introduced in \cite{Pu:2013:WGR:2500423.2500436} using transmit and receive arrays/antennas along with Fourier/Doppler analysis of the RSS data. They achieved a recognition accuracy of $94\%$ to classify 9 whole-body gestures in a home environment. Similar custom hardware-based solutions (e.g. \cite{AllSee}) are limited to the application environments and the resources. Nevertheless, antenna-array based methods find sophisticated novel solutions like seeing through walls using radio signals \cite{wi-vi}.

In \cite{DBLP:journals/corr/NandakumarKG14}, the authors used both the Wi-Fi RSS and the Channel State Information (CSI) to recognize hand gestures using a signal conditioning and thresholding based gesture recognition algorithm, and achieved a classification accuracy of $91\%$ on a laptop. 
Note, the CSI provides detailed channel features including the sub-carrier level phase information, but is supported only by a very limited set of Wi-Fi devices\footnote{Currently, the CSI data is available only from Intel's Wi-Fi Link 5300 drivers, and not available on smartphones Wi-Fi devices/drivers.}. Several other works such as \cite{al2016wiger,ali2015keystroke,WiG,zhang2016mudra} also proposed CSI-based solutions for gesture or activity recognition, however they are subjected to the same limitations.

K-Nearest Neighbors (K-NN) based classifiers have been widely used to recognize gestures from the RSS data. In \cite{sigg6813955} and \cite{christoph7134018}, the authors used window-based statistical features (e.g. mean, variance, maximum, number of peaks, etc.) applied to a K-NN classifier for recognizing hand gestures on a smartphone. They achieved accuracies of $50\%$ (with K=20, 11 hand gestures) and $90\%$ (with K=5, four hand gestures) respectively. 
However, their solutions require modified device firmware, root access to the OS, and dedicated applications that limits smartphone's Wi-Fi traffic.

A Discrete Wavelet Transform (DWT) based approach is presented in \cite{WiGest2015} to transform the RSS data intro three primitive signals: rising edges, falling edges and pauses. It achieved high accuracy ($90\%$) by using a classifier that compares the series of primitive signals to a set of pre-defined rules. However, such a solution requires high-frequency sampling of RSS and extensive computation abilities.

We depart from the literature works in four different ways: (1) we do not make any modifications to the existing hardware or software applications of the phone; (2) we introduce a new traffic induction approach to enable high-frequency RSS measurements; (3) we use custom but simple signal processing techniques and an efficient LSTM-RNN machine learning method to classify over-the-air hand gestures; (4) we share the experiment datasets and the partial source codes of our solution for the community to build this research further.


To the best of our knowledge, very few works uses deep learning or neural network based methods to radio signal based activity/gesture classification. For example, the authors in \cite{lv2017qualitative} used a Convolutional Neural Network (CNN) for classifying user driving behaviors based on narrow-band radio signals and achieved $88\%$ accuracy. On a higher level, the CNN is ideal for image processing due to its nature of recognizing patterns within the data (across space), whereas the RNN is ideal for speech/time-series processing due to its between-data (across sequence) recognition capabilities. Thus we choose RNN as the core machine learning algorithm in our Wisture solution.

\section{Background}
\subsection{Radio signal propagation}\label{sec:propagation}
An RF signal propagating through a medium is subject to several environmental factors that impact its characteristics. In the absence of nearby obstacles, the signal strength will be reduced by the free-space path loss (FSPL) caused by the spreading out of the signal energy in space. 
The RSS is usually modeled using a log-distance path loss model (an extension of Friss transmission equation) \cite{lindhe2013exploiting,1697062}:
\begin{equation}
{P_{r(dBm)}} = \underbrace {L_0 - 10n \log_{10} \left(R\right)}_{\text{Path Loss}} + \underbrace {\chi_{\sigma}}_{\text{Shadowing}} + \underbrace \varepsilon _{\text{multipath}},
\label{eq:6-13}
\end{equation}
\noindent where $L_0$ is the reference RSS at 1 meter from the transmitter (which depends on transmitter and receiver antenna properties as well as the signal frequency), $n$ is the decay exponent depending on the environment, $R$ is the distance between the transmitter and receiver, $\chi_{sigma} = \mathcal{N}(0,\sigma)$ is a zero-mean Gaussian variable with variance $\sigma^2$ used to represent the shadow fading caused by surrounding objects due to reflection, absorption, and diffractions~\cite{Ghasemi2012}. $\varepsilon$ is a random variable following the Nakagami distribution  used to represent multi-path fading caused by the fact that the signals take multiple paths to the destination, and to represent interferences and spatio-temporal noise~\cite{lindhe2013exploiting}. 
The combination of $\chi_{\sigma}$ and $\varepsilon$ models the different ways by which human body presence and movements impacts the strength of the radio signal received by a nearby wireless device, such as a smartphone.

\begin{figure}[t]
\centering
\includegraphics[width=\linewidth]{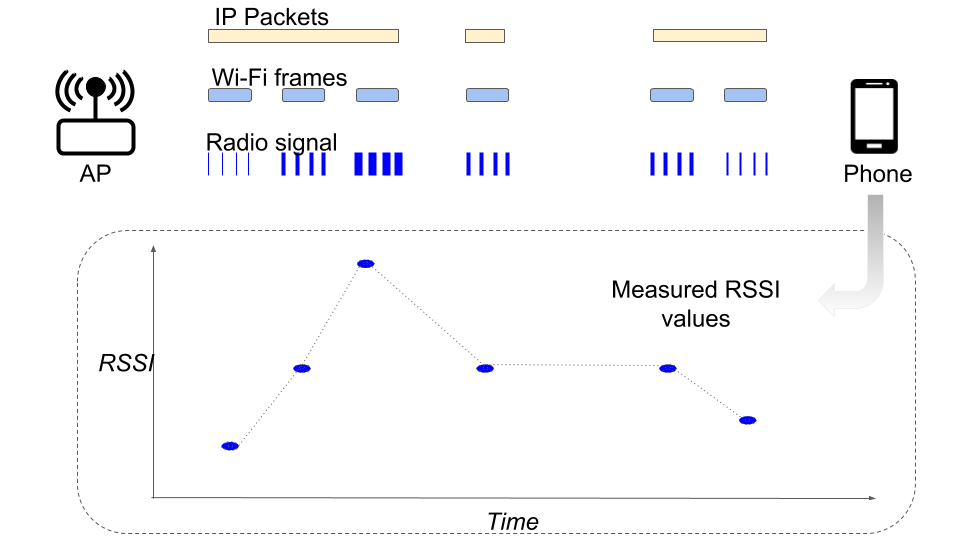}
\caption{Illustration of RSSI measurements in a smartphone.}
\label{fig:rssi}
\end{figure}
\begin{figure*}[t]
\centering
\includegraphics[width=0.32\textwidth]{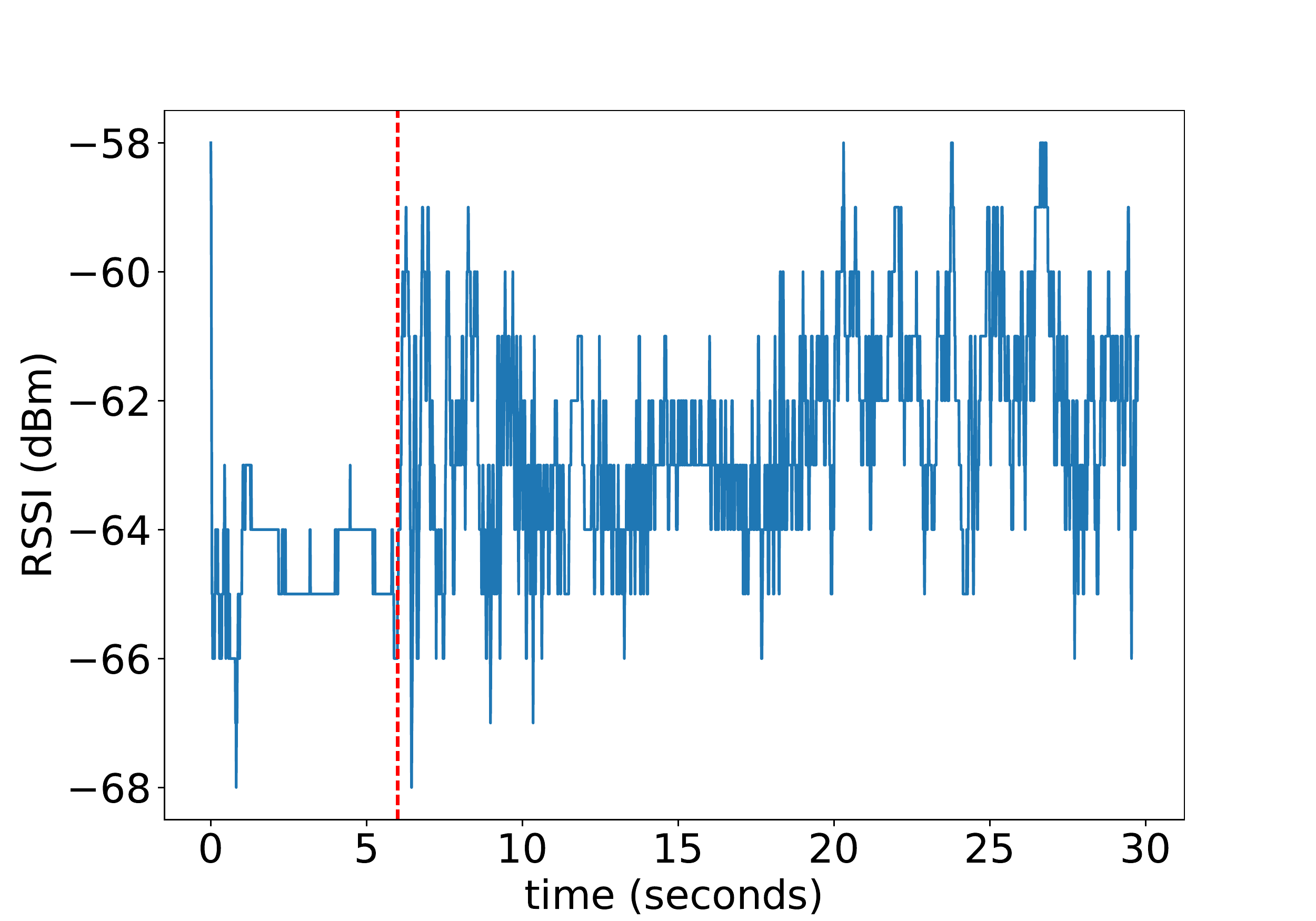}
\includegraphics[width=0.32\textwidth]{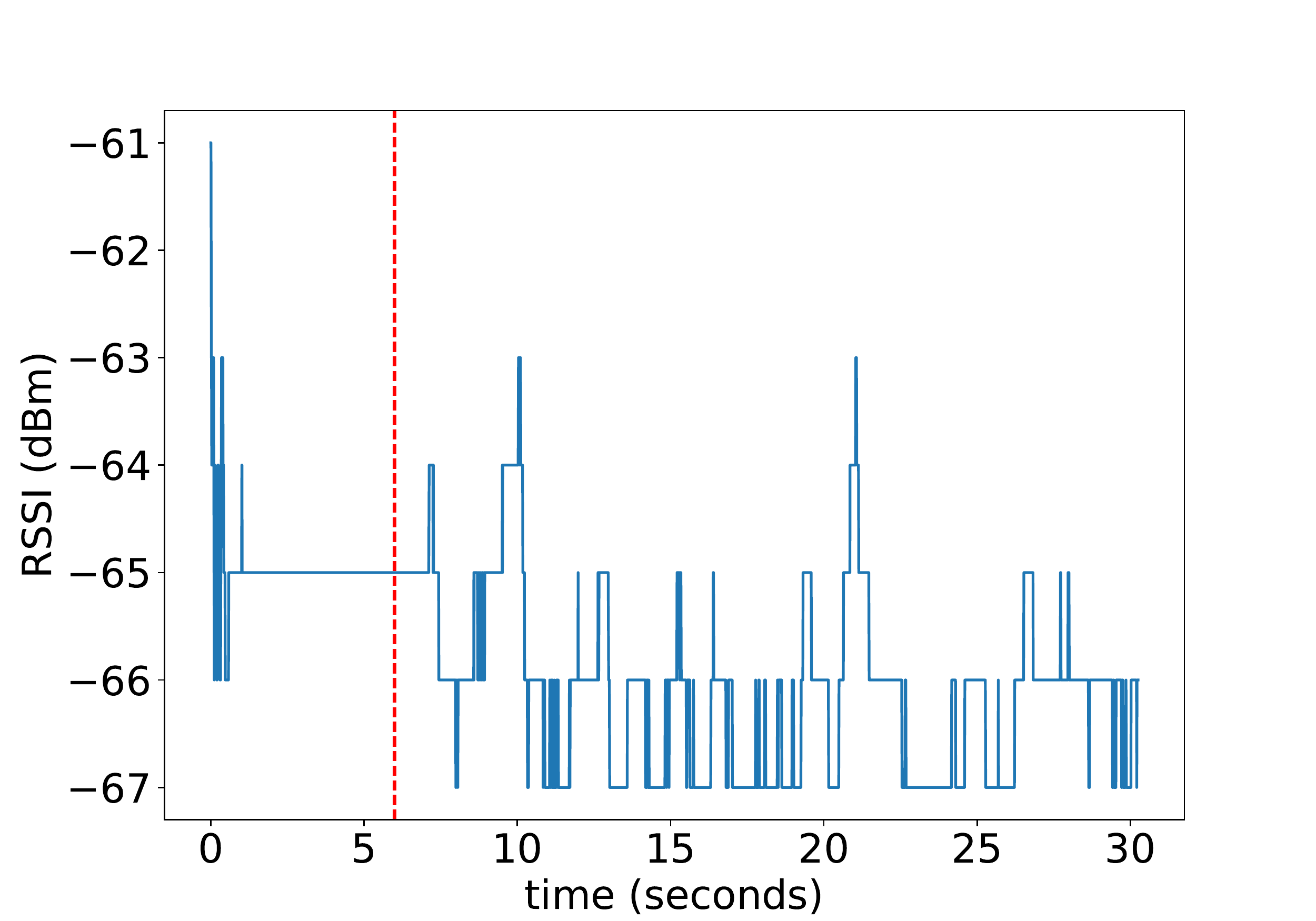}
\includegraphics[width=0.32\textwidth]{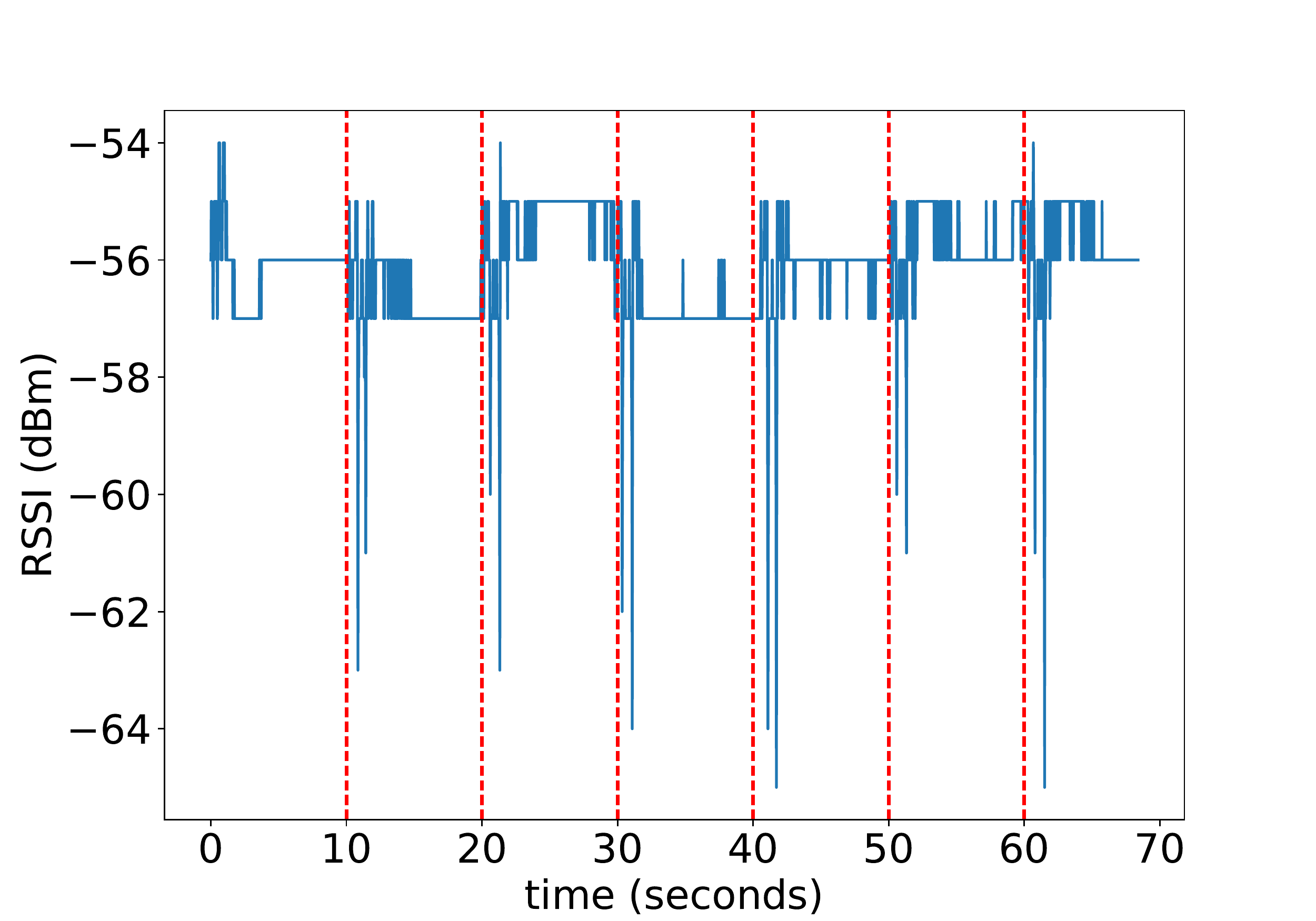}
\caption{RSSI measurements of a smartphone placed on table near a person performing different activities in a room: walking past (left); typing on a keyboard (middle); and making swiping gestures every 10 seconds (right). The red marker indicates when the person started walking, typing or swiping the hand. The AP is placed about 2 meters from the smartphone.}
\label{fig:activities}
\end{figure*}
\subsection{Wi-Fi RSS measurements}\label{sec:rssi}
Wi-Fi technology is based on the IEEE 802.11 standards and typically use 2.4 or 5 GHz ISM frequency spectrum. The data is communicated as frames. For every received frame, the device measures the RSS and reports the Received Signal Strength Indicator (RSSI), usually in dBm. 
Figure~\ref{fig:rssi} shows an illustration of a sequence of RSSI measurements generated by a Wi-Fi receiver.
The RSSI together with additional measurements (e.g Link Quality, Noise level, etc.) at the interface provide information on the transmission channel. 
The frequency by which the RSSI measurements are taken is not deterministic, but depends on the traffic sent to the Wi-Fi receiver. This burstiness nature of the RSSI measurements makes it hard to use for gesture detection. 

In an idle Wi-Fi network, the majority of the traffic is formed of so-called beacon frames. Beacon frames are sent periodically by the AP to signal the presence of the Wi-Fi network. The time between two beacon frame transmissions is configurable, but typically set to 102 milliseconds \cite{jim}. This means that a smartphone connected to a Wi-Fi network, and not actively receiving data, will have around nine RSSI measurements every second corresponding to the beacon frames received during that second.

Note the RSSI measurements made on a laptop or a computer with commercial Wi-Fi adapters, can provide much stronger and high-frequency data compared to measurements on a smartphone. This is likely due to the antenna properties, device drivers, and limited resources on a smartphone, which alludes to the difficulty of using Wi-Fi RSSI for recognizing gestures in smartphones compared to computers.

\subsection{Hand gestures}\label{sec:gesture_desc}
We first show in Fig.~\ref{fig:activities} the RSSI data recorded in a smartphone while a person is performing an activity such as walking past the phone, typing on a keyboard, and swiping his/her hand over the phone. Note the unique pattern in the RSSI stream created by the hand gesture.

In this paper, we consider three hand gestures, see Fig.~\ref{fig:gestures}: swipe, push, and pull. 
\begin{itemize}
\item \textit{Swipe gesture}:  moving the hand about five centimeters above the smartphone from one side to the other and back to the starting point.
\item \textit{Push gesture}: moving the hand downward towards the smartphone and holding it steadily about five centimeters above it for around two seconds.
\item \textit{Pull gesture}: placing the hand about five centimeters above the smartphone, holding it  there for about two seconds before moving it upward.
\end{itemize}
If more gestures are needed one could either introduce new gestures, or make combinations of these three primitive gestures as in \cite{WiGest2015}.
The question of which option is best is however beyond the scope of this paper, as we focus on the best machine learning solution.

\section{Problem formulation}
The problem of recognizing hand gestures from RSS values can be viewed as a classification problem where the objective is to learn a mapping from the RSS values to the probability distribution over the possible hand gestures $P(y|x)$.
\begin{equation}
  x \to P(y|x;\theta)
\end{equation}
Where $x \in \mathbb{R}^{\tau}$ is a sequence of input RSS values with length $\tau$, $y\in\{0,1,\ldots,K\}$ is the list of gestures, $\theta$ is a parameterization of the mapping.

Given a dataset of $m$ sample gestures that is formed from the inputs $X=[x_1\ x_2 \ldots x_m]$ and their corresponding outputs $Y=[y_1\ y_2 \ldots y_m]$. The maximum likelihood (ML) method can then be used to find a good estimate of $\theta$ as below:
\begin{equation}
\label{eq:ml}
  \theta_{ML}=\operatorname*{arg\,max}_{\theta} P(Y|X;\theta)
\end{equation}

Assuming the dataset samples are independent and identically distributed (i.i.d.). Equation~\ref{eq:ml} can be rewritten as follows:
\begin{equation}
  \theta_{ML}=\operatorname*{arg\,max}_{\theta} \prod\limits_{i=1}^m P(y_i|x_i;\theta)
\end{equation}
The above probability product can become very small and hence render the problem computationally unstable. This can be solved by taking the logarithm of the likelihood, which transforms the product of probabilities into a sum.
\begin{equation}
  \theta_{ML}=\operatorname*{arg\,max}_{\theta} \sum\limits_{i=1}^m \log P(y_i|x_i;\theta)
\end{equation}
The estimate of $\theta$ can be expressed as minimizing a loss function $L$ (also referred to as cost) defined as below 
\begin{equation}
L = \sum\limits_{i=1}^m -\log P(y_i|x_i;\theta)
\end{equation}
This loss is known as \textit{negative log-likelihood (NLL)}.

As will be described below, an RNN is used to model the conditional probability $P(y|x;\theta)$. An ML estimate of $\theta$ is found  using the Stochastic Gradient Descent (SGD) algorithm, to minimize the NLL in a collected hand gesture dataset.

\begin{figure*}[t]
\centering
\includegraphics[scale=0.65]{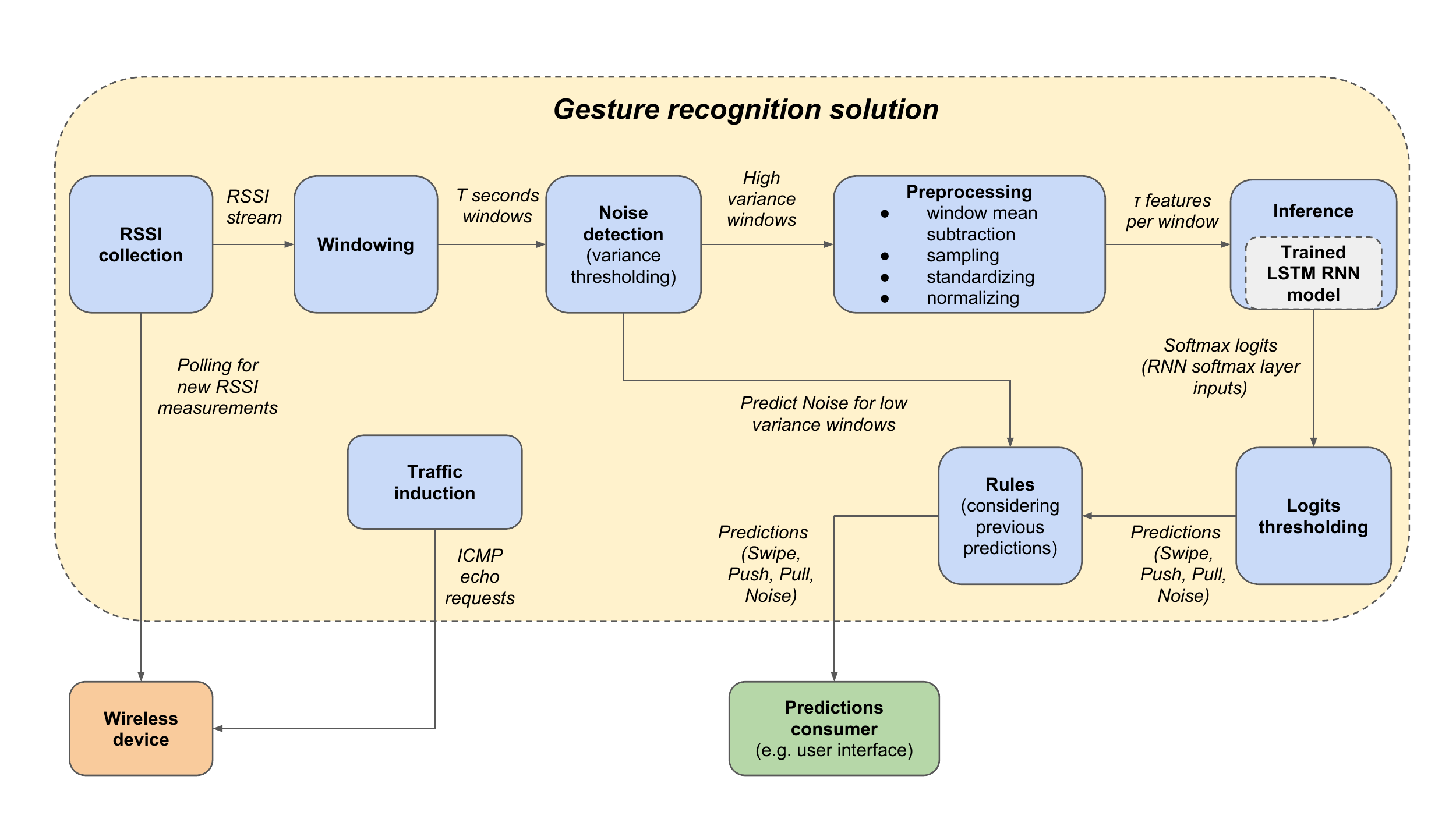}
\caption{Wisture: LSTM RNN based gesture recognition using Wi-Fi RSS in smartphones.}
\label{fig:endend}
\end{figure*}
\section{Proposed solution: Wisture}\label{sec:sol_overview}
We first describe the Wisture solution in general and then present how to train the system with the training data.
\subsection{Signal processing and machine learning method}
Figure~\ref{fig:endend} shows an overview of the proposed gesture recognition solution. The different submodules are described below.
\subsubsection{Traffic induction}\label{sec:induction}
As discussed earlier, the smartphone Wi-Fi interface makes new RSSI measurement only when a new Wi-Fi frame is received. To guarantee that the wireless device makes enough updated RSSI measurements, we induce artificial traffic between the AP and the smartphone by sending a continuous stream of Internet Control Message Protocol (ICMP) echo requests to the AP. For every ICMP echo request, the AP will send an ICMP echo reply back to the smartphone which will make an updated RSSI measurement. 
This enables us to have enough RSSI measurements while avoiding the need for a custom firmware or putting the Wi-Fi interface in monitor mode as in~\cite{christoph2015,sigg6813955}.

\subsubsection{RSSI collection}
This module extracts a stream of up to $\sim200$ RSSI values/second from the Wi-Fi interface. In our Android implementation, the RSSI measurements are collected using the \textit{wireless extension for Linux} user interface~\cite{linux}, which is exposed as a pseudo file named \textit{/proc/net/wireless}.

\subsubsection{Windowing}\label{sec:windowing}
The incoming RSSI stream is split into overlapping windows of $T$ seconds length, and $d$ seconds gaps between window starts. In all the experiments the gap $d$ was set to 1 second. Different values of $T$ are investigated and reported in the experiments section. Since the incoming RSSI stream rate is varying around $200$ values per second, the output windows will have a variable length.

\subsubsection{Noise detection}
We note that only windows with high activity, identified by a theshold on the window variance, are likely to be caused by hand gestures. Thus only the windows that pass this criteria will be forwarded to the subsequent steps. All windows that have a variance less than the threshold will be predicted as \textit{no gesture} or \textit{Noise}. The process of estimating the variance threshold is described in Sec.~\ref{sec:thresh_estim}.

\subsubsection{Preprocessing}\label{sec:preprocess}
This submodule transforms the incoming windows into windows of equal number of feature values ($\tau$). Each incoming window is processed as below:
\begin{itemize}
\item \textit{Mean subtraction}: The window values are centered around zero, by subtracting the window mean from all RSSI values in the window. This increases the system robustness against changes in the RSSI values due to, for example, RSSI increases or decreases when the smartphone is moved close or away from the AP respectively.
\item \textit{Sampling}: This steps samples $\tau$ feature values with a time difference between consecutive samples equal to $T/\tau$ on average. Different values of $\tau$ have been investigated and reported in the experiments section.
\item \textit{Standardizing}: Each one of the $\tau$ feature values is reduced by the training data mean of that feature value. 
\item \textit{Normalizing}: Each one of the $\tau$ feature values (standardized in the previous step) is divided by the training data standard deviation of that feature value.
\end{itemize}
A detailed description of the standardizing and normalizing steps is provided in Sec.~\ref{sec:offpreprocess}.

\subsubsection{Inference}
The LSTM RNN model takes an input of $\tau$ feature values, and outputs three values proportional to the conditional probability of each possible gesture on the inputs. These values are referred to as \textit{logits}, because they are the inputs of the \textit{softmax} layer used in calculating the model loss during training. The Softmax function is a generalization of the logistic function, and its inputs are referred to as logits.

\subsubsection{Logits thresholding}
This step discards LSTM RNN model predictions that are below a specific threshold and predicts \textit{Noise} for those inputs.

\subsubsection{Prediction decision rules}\label{sec:rules}
\begin{figure*}[h]
  \centering
  \includegraphics[width=0.32\textwidth]{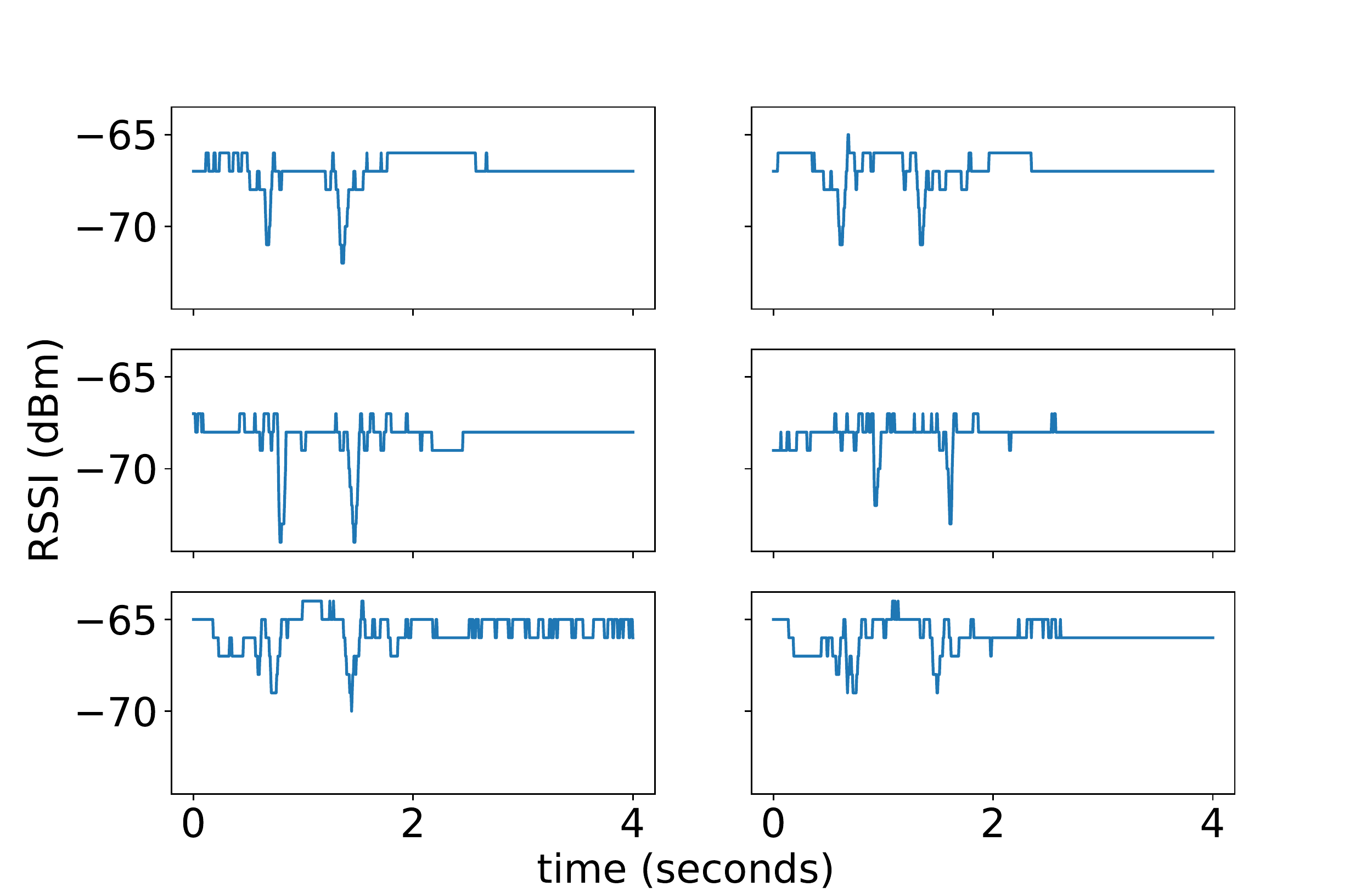}
  \includegraphics[width=0.32\textwidth]{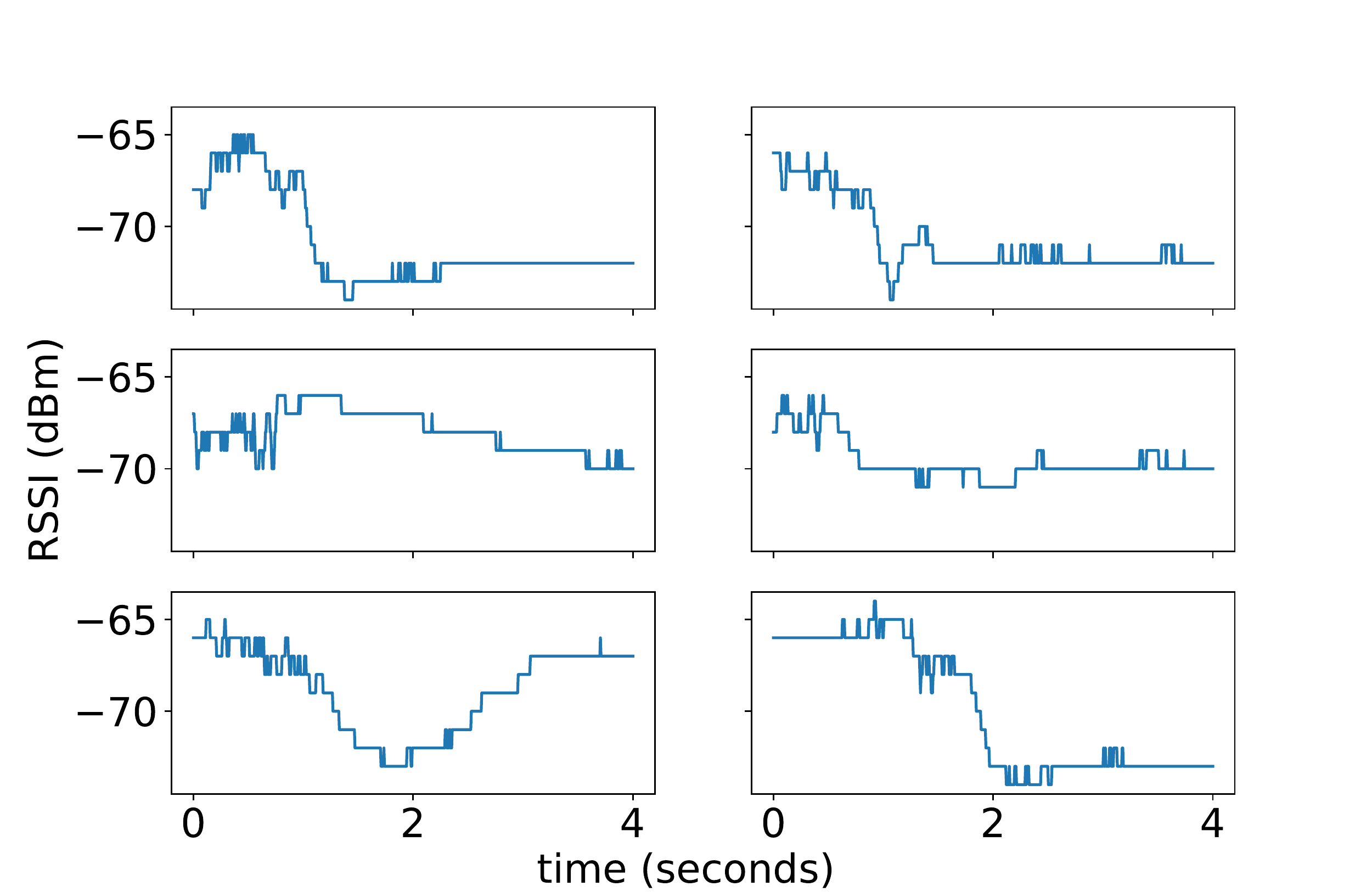}
  \includegraphics[width=0.32\textwidth]{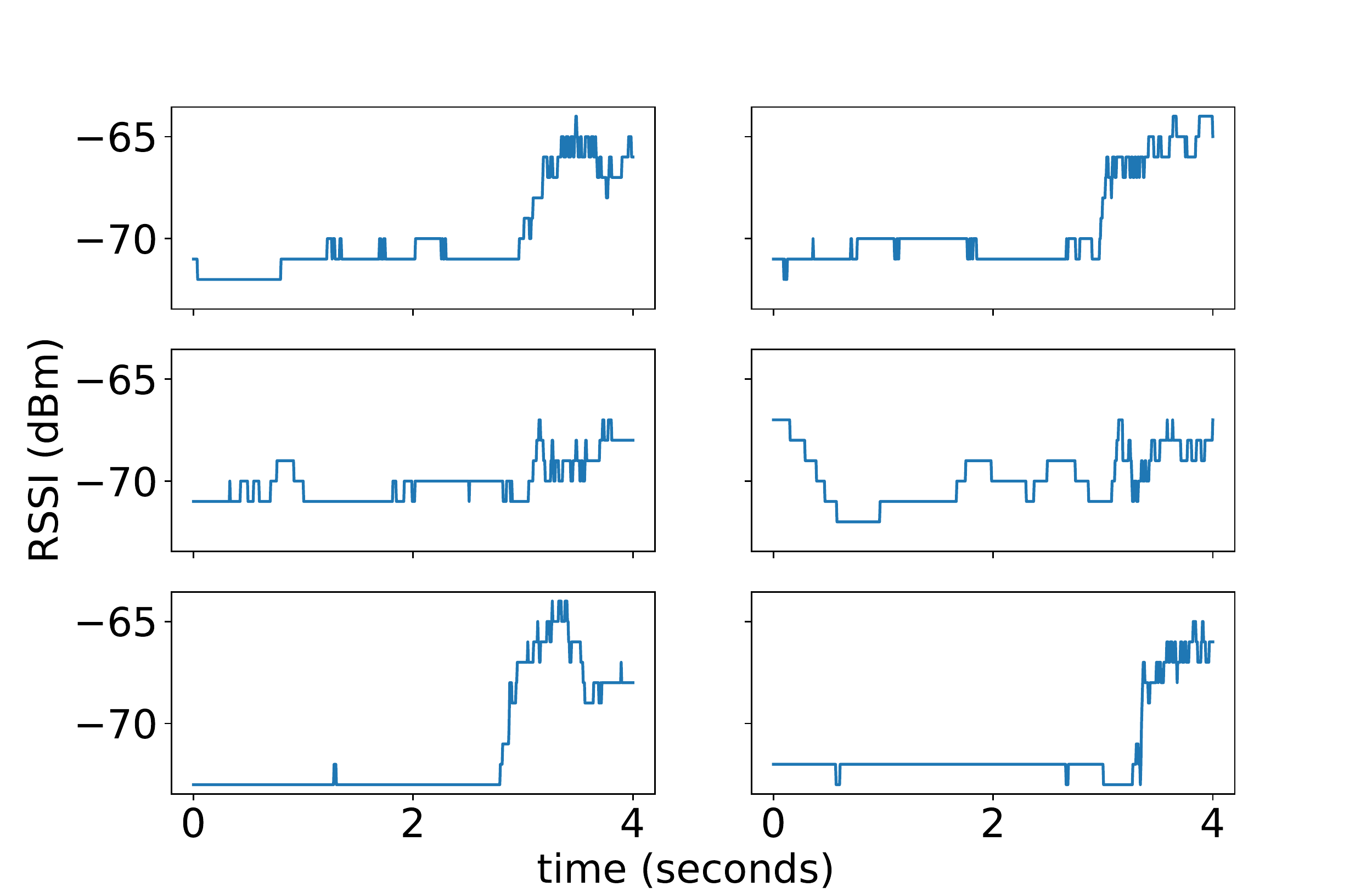}
  \caption{Sample Gesture windows: Swipe (left); Push (middle); Pull (right).}
  \label{fig:sample_gestures}
\end{figure*}

This submodule keeps a short history of the previous predictions, and applies a set of rules to accept or reject the current prediction made by the preceding steps. These rules are:
\begin{itemize}
\item Allow Pull gestures only after Push gestures. The Pull gesture RSSI signature appears as a pause followed by an increase. This signature is similar to those caused by some background activities, e.g. when an AP increases its output signal power. This rule reduces the number of false positive Pull predictions caused by such interfering background activities\footnotemark. 

\footnotetext{Note: However, the solution is prone to confusing decreases in RSSI values caused by interfering background activities (e.g. the AP reducing its output signal power) as Push gesture, and no solution was proposed in this work to harden the system against such interference.}
\item A prediction that is different from its immediate predecessor is ignored (and Noise is predicted instead). Exempted from this rule are: 
\begin{itemize}
\item Swipe or Push following a Noise prediction.
\item Pull prediction that follows a Push.
\item Noise predictions.
\end{itemize}
The rationale for this rule is that each prediction window overlaps with the previous window (three seconds overlap in most experiments). In many cases, if the preceding window contained a gesture, the succeeding window RSSI stream signature might look similar to another gesture than the performed one. For example, the end of Swipe gestures look similar to Pull gestures (see Fig.~\ref{fig:sample_gestures}).
\end{itemize}

\subsection{System training}\label{sec:training}
Training the LSTM RNN model and selecting the different hyperparameters and thresholds are performed in an offline setting. The training procedure is detailed below.

\subsubsection{Data preprocessing}\label{sec:offpreprocess}
The online machine learning solution preprocesses the incoming RSSI windows in the way described below.
\begin{itemize}
\item The RSSI values are read from the collected data files, and then split into $D$ windows (corresponding to the gestures), each being $T$ seconds long.
\item For each window, the mean is calculated and then subtracted from the individual window values.
\item $\tau$ values that are equally spaced in time are then sampled from each window. The result is a dataset of shape $D$ windows each having $\tau$ features.
\item The dataset is then randomly split into training ($D_{train}=0.75D$) and testing ($D_{test}=0.25D$) sets. Furthermore, when a model hyper parameter selection is done, $0.8D_{train}$ of the training set is used to train the model, and the remaining $D_{val}=0.20D_{train}$ is used to select the hyper parameters (validation set).
\item Using the training set ($D_{train}\times \tau$), the mean and standard deviation of each one of the $\tau$ features is calculated as below.
\begin{equation}
x^{(i)}_{train\_mean}=\frac{1}{M}\sum_{j=1}^{M} x^{(i)}_{train, j}
\end{equation}
\begin{equation}
x^{(i)}_{train\_std}=\sqrt{\frac{1}{M}\sum_{j=1}^{M} (x^{(i)}_{train, j} - x^{(i)}_{train\_mean})^2}
\end{equation}
Where $M$ is the training set $D_{train}$ size and $x^{(i)}_{train, j}$ is feature $i$ value of sample $j$ from the training set.
\item All training and testing set windows were standardized and normalized using the training mean and standard deviation. Let $x=[x^{(1)}\ x^{(2)} \ldots x^{(i)} \ldots x^{(\tau)}]$ be some input window (from training or testing), the output of the standardization and normalization steps $x_o=[x^{(1)}_o\ x^{(2)}_o \ldots x^{(i)}_o \ldots x^{(\tau)}_o]$ can be described as below:
\begin{equation}
x^{(i)}_{o}= \frac{x^{(i)}- x^{(i)}_{train\_mean}}{x^{(i)}_{train\_std}}
\end{equation}
\end{itemize}

\subsubsection{LSTM RNN model training}\label{sec:lstm_training}
\begin{figure}
\centering
\includegraphics[width=0.5\textwidth]{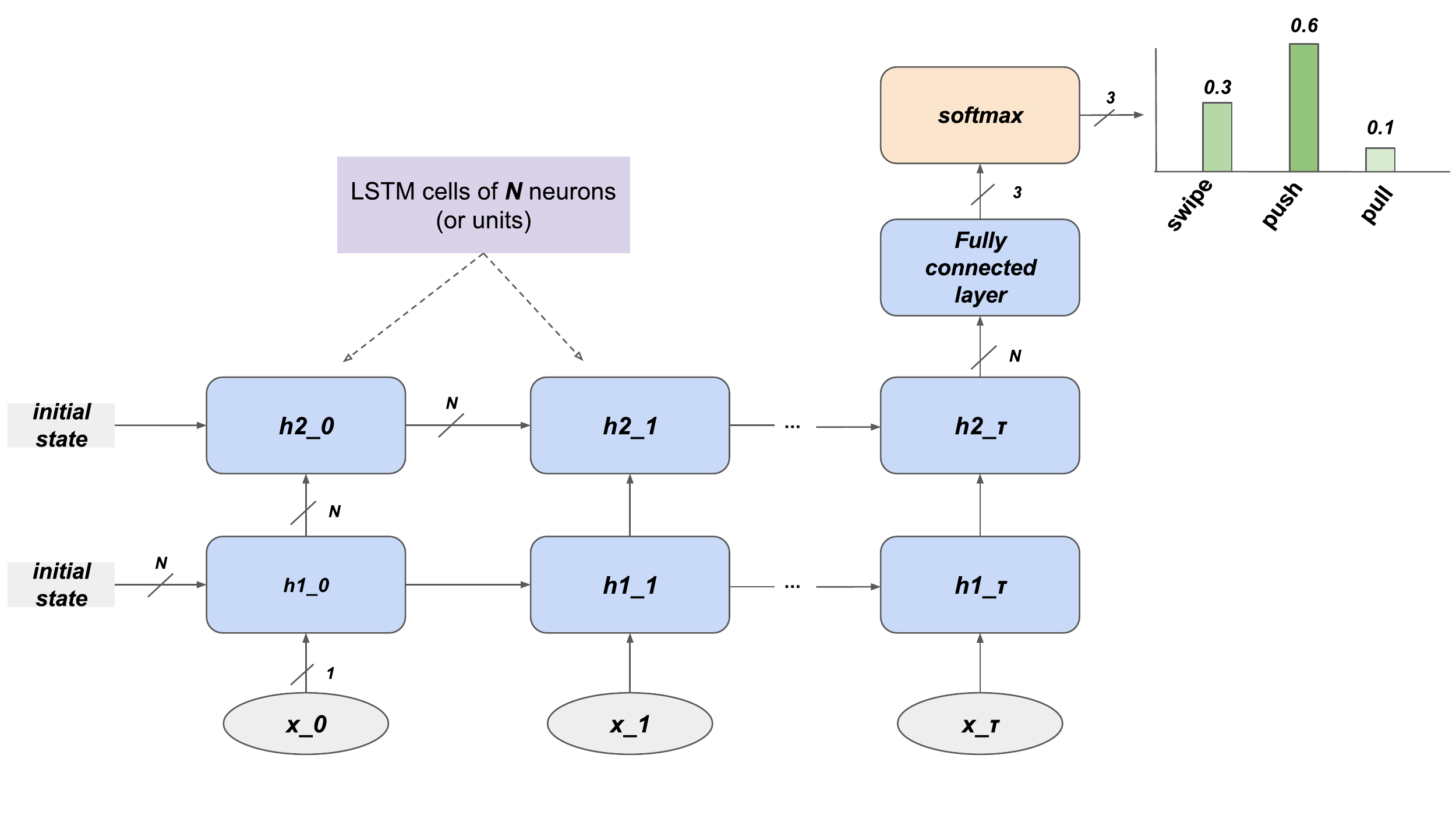}
\caption{Illustration of the (time-unrolled) LSTM RNN model used in \textit{Wisture}.}
\label{fig:model}
\end{figure}

Figure~\ref{fig:model} shows an illustration of the LSTM RNN model used in this work (inspired by \cite{DBLP:journals/corr/ZarembaSV14}). The model was trained to minimize the NLL loss, using a variant of SGD known as \textit{Adaptive Moment Estimation}, or shortly \textit{ADAM}.
Most of the model hyper parameters ($\tau$, $N$, number of layers and others) were selected by performing a grid search in the parameters space. Each parameter setting is evaluated using a four folds cross validation.
\subsubsection{Thresholds selection}\label{sec:thresh_estim}
The variance threshold used by the Noise detection step, is initially estimated as the training data minimum  window variance. This value is then manually optimized to maximize the online prediction accuracy. The same approach is used to select the logits thresholds.

\section{Experiments and Results}
\subsection{Experiment Dataset}
This section contains the details of the dataset used to train and evaluate the recognition system.

\subsubsection{Spacial setup}
\begin{figure}[b]
  \centering
  \includegraphics[width=1\linewidth]{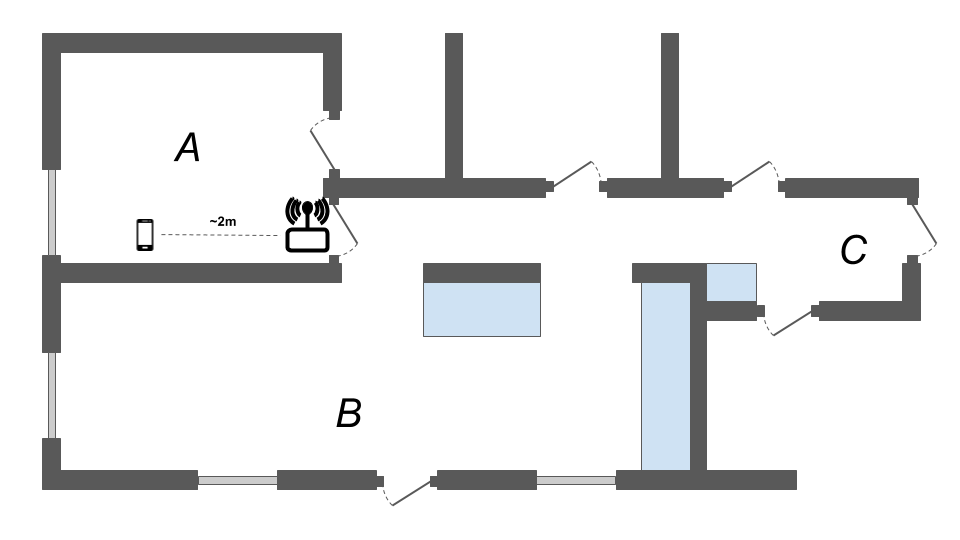}
  \caption{Floor plan of the experiment set-up used in data collection and system testing.}
  \label{fig:spatial_setup}
\end{figure}

Figure~\ref{fig:spatial_setup} presents the experimental setup. The dataset was collected under two different spatial configurations of the Wi-Fi AP and the smartphone:
\begin{itemize}
\item (Room A): The AP and the smartphone were placed two meters apart in room $A$ with a line of sight (LoS) between them. The AP was placed on a table slightly lower than the table where the smartphone was placed.
\item (Room A \& B): The AP was placed in room $B$ and the smartphone in room $A$, see Fig.~\ref{fig:spatial_setup}, and the distance between them was $\sim$4.5 meters. The two rooms were separated by a wall made mainly of wood and gypsum, thus there is no line of sight (nLoS). Both the AP and the smartphone were placed on tables of similar height. 
\end{itemize}

\subsubsection{Traffic scenarios}
Three different WiFi traffic scenarios were considered when collecting the data:
\begin{itemize}
\item (Internet access + traffic induction): in this scenario the AP was connected to the Internet and  the smartphone was connected to the AP. The smartphone was continuously sending ICMP requests to the AP (pinging) at a rate of $\sim$700 times/second.
\item (No Internet access + traffic induction): neither the AP nor the smartphone had Internet access, but the smartphone was continuously pinging the AP at a rate of $\sim$700 times/second.
\item (No Internet access + no traffic induction): neither the AP nor the smartphone had Internet access, and there was no traffic induction. 
\end{itemize}

\subsubsection{Data collection procedure}\label{sec:collection}
An Android mobile application was developed specifically for recording the Wi-Fi RSSI data using the induced traffic approach (called "Winiff"). It records the RSSI measurements made by the smartphone at a frequency of $\sim$200 samples/second.
A typical collection session is described below:
\begin{enumerate}
\item The AP and the smartphone are placed as per one of the spatial setups described earlier.
\item The subject performing the experiment sits in a chair facing the smartphone.
\item The smartphone is connected to the AP.
\item The RSSI collection application is started.
\item At a specific point in time (\textit{start time}), the subject starts performing the gestures, leaving a gap of ten seconds between consecutive gestures (\textit{gap time}). Both the start and gap times are noted down and used later to extract the gesture windows.
\item The collected RSSI stream is stored in a text file.
\end{enumerate}
The collected dataset details are summarized in Table~\ref{table:data}.
\begin{table}
  \centering
  \begin{tabular}{l l l l l}
    \hline
    Name & Location & Induction & Internet & \#Samples \\
    \hline
    Dataset1 & room A & $\surd$ & & 440 \\
    Dataset2 & room A & $\surd$ & $\surd$ & 432 \\
    Dataset3 & room A &  & & 434\\
    Dataset4 & room A \& B & $\surd$ & & 337\\
    \hline
  \end{tabular}
  \caption{Summary of the collected datasets and the total number of gesture samples per dataset.}
  \label{table:data}
\end{table}

\subsubsection{Evaluation metric}
The system was evaluated using the \textit{accuracy} measure, defined as the percentage of correctly predicted gestures. For a set of test gestures $X=[x_1\ x_2\ldots x_i \ldots x_m]$ with corresponding true labels $Y=[y_1\ y_2 \ldots y_i \ldots y_m]$, the accuracy is defined as:
\begin{equation}
accuracy = 100 \times \frac{1}{m}\sum\limits_{i=1}^m I_{y_i}(\hat{y_i}) \;\;\;  \%
\end{equation} 
Where $\hat{y_i}$ is the system prediction for input gesture $x_i$, and $I_{y_i}(\hat{y_i})$ is $1$ if $\hat{y_i} = y_i$ and $0$ otherwise.

\subsection{Results and Discussions}
Here we list the conducted experiments and the obtained results. The RNN model was trained as described in Sec.~\ref{sec:lstm_training}. The model parameters used in the experiments are shown in Table~\ref{table:params}.
The reported mean accuracies in  experiments are calculated by evaluating the RNN model ten times on the specific configurations being tested, each using a different random split of the data into training and testing sets.
\begin{table}
  \centering       
  \begin{tabular}{l l}
    \hline
    Parameter & Value \\
    \hline
    \multicolumn{1}{l}{RNN time steps (T)} & 50 \\
    \multicolumn{1}{l}{Number of hidden LSTM layers} & 2 \\
    \multicolumn{1}{l}{Number of units (or neurons) per LSTM (N)} & 200 \\
    \multicolumn{1}{l}{Learning rate} & 0.001 \\
    \multicolumn{1}{l}{SGD batch size} & 50 \\
    \multicolumn{1}{l}{Dropout probability} & 0.5 \\
    \multicolumn{1}{l}{Parameters initial random values boundaries} & $\pm$0.08 \\
    \multicolumn{1}{l}{Maximum gradient norm (for clipping gradients)} & 25 \\
    \multicolumn{1}{l}{Number of training iterations} & 600 \\
    \hline
  \end{tabular}
  \caption{LSTM RNN parameters and hyper parameters.}
  \label{table:params}
\end{table}

\subsubsection{RNN model accuracy on different datasets}\label{sec:induction_impact}
\begin{figure}
\centering
\vspace{-6mm}
\includegraphics[width=0.8\columnwidth,height=0.55\columnwidth]{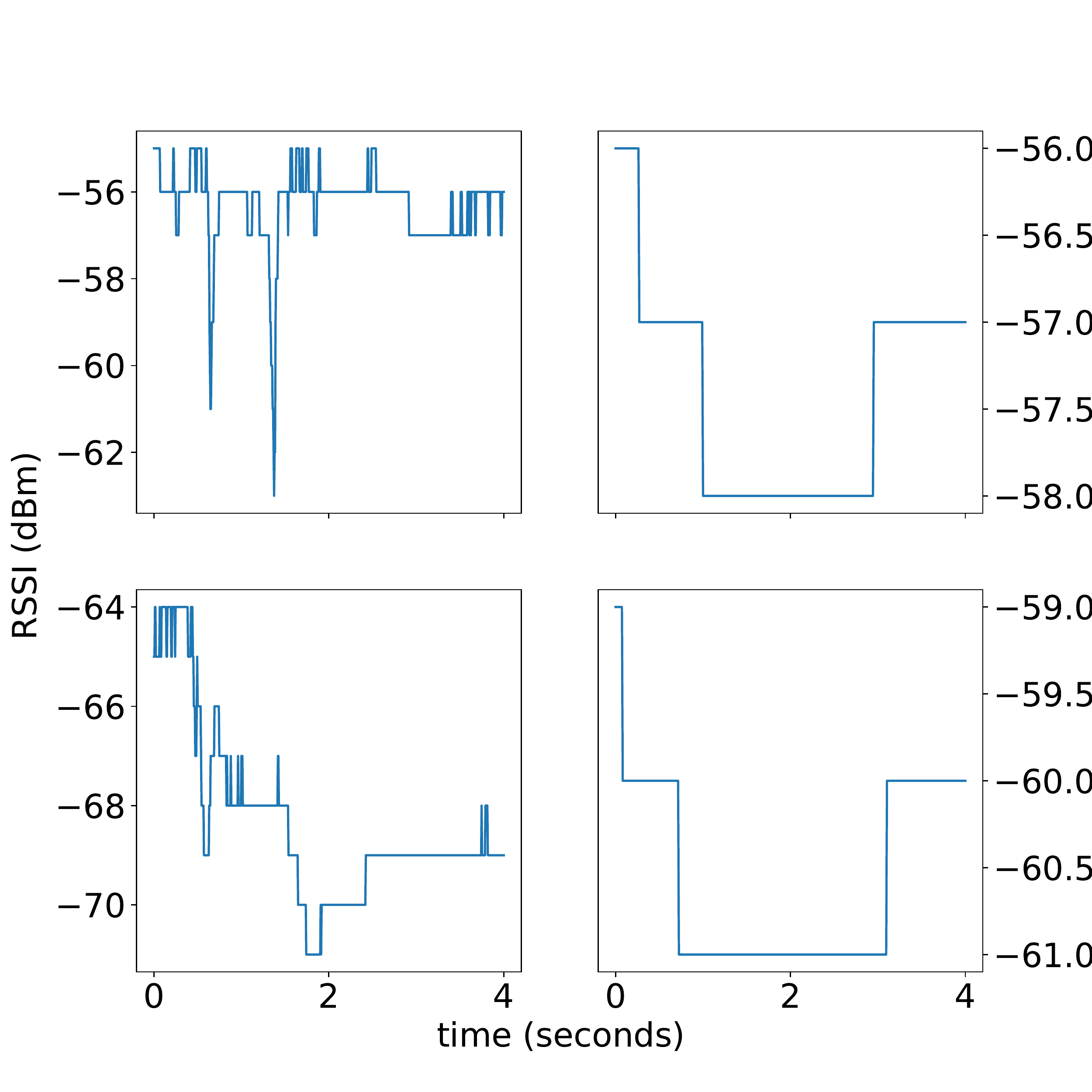}
  \caption{The left column shows the RSSI values during Swipe (top) and Push (Bottom) gestures with the traffic induction enabled. The right column shows the corresponding gestures with the traffic induction disabled.}\label{fig:impact_no_traffic}
\end{figure}

Table~\ref{table:offline_ds_res} lists the recognition model accuracy when evaluated on the different datasets. 
Due to the increased data size of the collective dataset (Dataset1 + Dataset2 + Dataset4), the number of training iterations for this configuration is 1000 instead of 600.
\begin{table}[h]
  \centering
  \begin{tabular}{l l}
    \hline
    Dataset Name & Accuracy ($\pm$std)\\
    \hline
    Dataset1 & $91\%$ ($\pm$3.1)\\
    Dataset2 & $83\%$ ($\pm$2.5)\\
    Dataset3 & $78\%$ ($\pm$2.4)\footnotemark  \\
    Dataset4 & $87\%$ ($\pm$2.9)\\
    Dataset1 + Dataset2 + Dataset4 & $94\%$ ($\pm$1.6) \\
    \hline
  \end{tabular}
  \caption{LSTM-RNN accuracy on the collected datasets.}
  \label{table:offline_ds_res}
\end{table}
\footnotetext{If we exclude the Swipe gesture in Dataset3, the prediction accuracy jumps to $97\%$ ($\pm$1.5).}
The gesture recognition accuracy is poor for Dataset3, which can be attributed to fact that only a small number of (low-frequency) RSSI measurements were available in the dataset because both the internet and the traffic induction were disabled. The significant increase in accuracy from $78\%$ to $91\%$ when the induction is enabled demonstrates the high dependency of the accuracy on the amount of data traffic at the Wi-Fi interface.
Figure~\ref{fig:impact_no_traffic} compares the RSSI values recorded for the Swipe and Push gestures performed while the induction is enabled and disabled. 
It is clear that with no traffic induction, the RSSI values corresponding to the Swipe gesture are not distinguishable from those of the Push gesture, and consequently the classification ability is severely impacted.

\subsubsection{Comparison with the state-of-the-art methods}
A set of time series classification algorithms, including state-of-the-art ones, were evaluated on Dataset1, to compare them to the RNN model (Table~\ref{table:all_accuracy}). The Collective of Transformation Ensembles (COTE), Elastic Ensemble (EE), Shapelet Transform ensemble (STE) \cite{DBLP:journals/corr/BagnallBLL16}, which are ensemble methods that employs multiple classifiers underneath, performed better than or equal to the RNN model. The Learning Time-series Shapelets (LTS)\footnote{We implemented a python version of the LTS method, available at \url{https://github.com/mohaseeb/shaplets-python}.} \cite{Grabocka:2014:LTS:2623330.2623613} performed better in terms of accuracy but poorer in terms of training/prediction time.. The K-Nearest Neighbor Dynamic Time Warp (K-NN DTW) algorithm and the Fast Shapelets (FS)~\cite{DBLP:journals/corr/BagnallBLL16} performed worse, with the k-NN DTW having a very slow prediction time of almost one second.
\begin{table}
  \centering
  \resizebox{\columnwidth}{!}{\begin{tabular}{l l l l}
    \hline
    \multicolumn{1}{l}{Algorithm} & Accuracy & Training time & Prediction time \\
    \hline
    \multicolumn{1}{l}{K-NN DTW} & $90\%$ ($\pm$28) & N/A & 964.15 ms\\
    \multicolumn{1}{l}{FS} & $85\%$ ($\pm$4.6) & 0.19 min & 0.01 ms\\
    \multicolumn{1}{l}{STE} & $91\%$ ($\pm$1.1) & 4.91 min& 26.86 ms\\
    \multicolumn{1}{l}{LTS} & $93\%$ ($\pm$2.3) & 19.09 min&  9.29 ms\\
    \multicolumn{1}{l}{EE} & $93\%$ ($\pm$1.7) & 10.95 min& 23.09 ms\\
    \multicolumn{1}{l}{COTE} & $94\%$ ($\pm$2.4) & 50.68 min& 178.20 ms\\
    \multicolumn{1}{l}{LSTM RNN} & $91\%$ ($\pm$3.1) & 3.29 min& 7.04 ms\\
    \hline
    \end{tabular}}
    \caption{Prediction accuracies, training and per-sample prediction times reported for various machine learning solutions evaluated on Dataset1. The Wisture solution had a balanced performance in terms of high accuracy and low prediction and training time.}
  \label{table:all_accuracy}
\end{table}

\subsubsection{Impact of dataset size on the accuracy}
\begin{figure}
\centering
\vspace{-4mm}
  \includegraphics[width=0.7\linewidth]{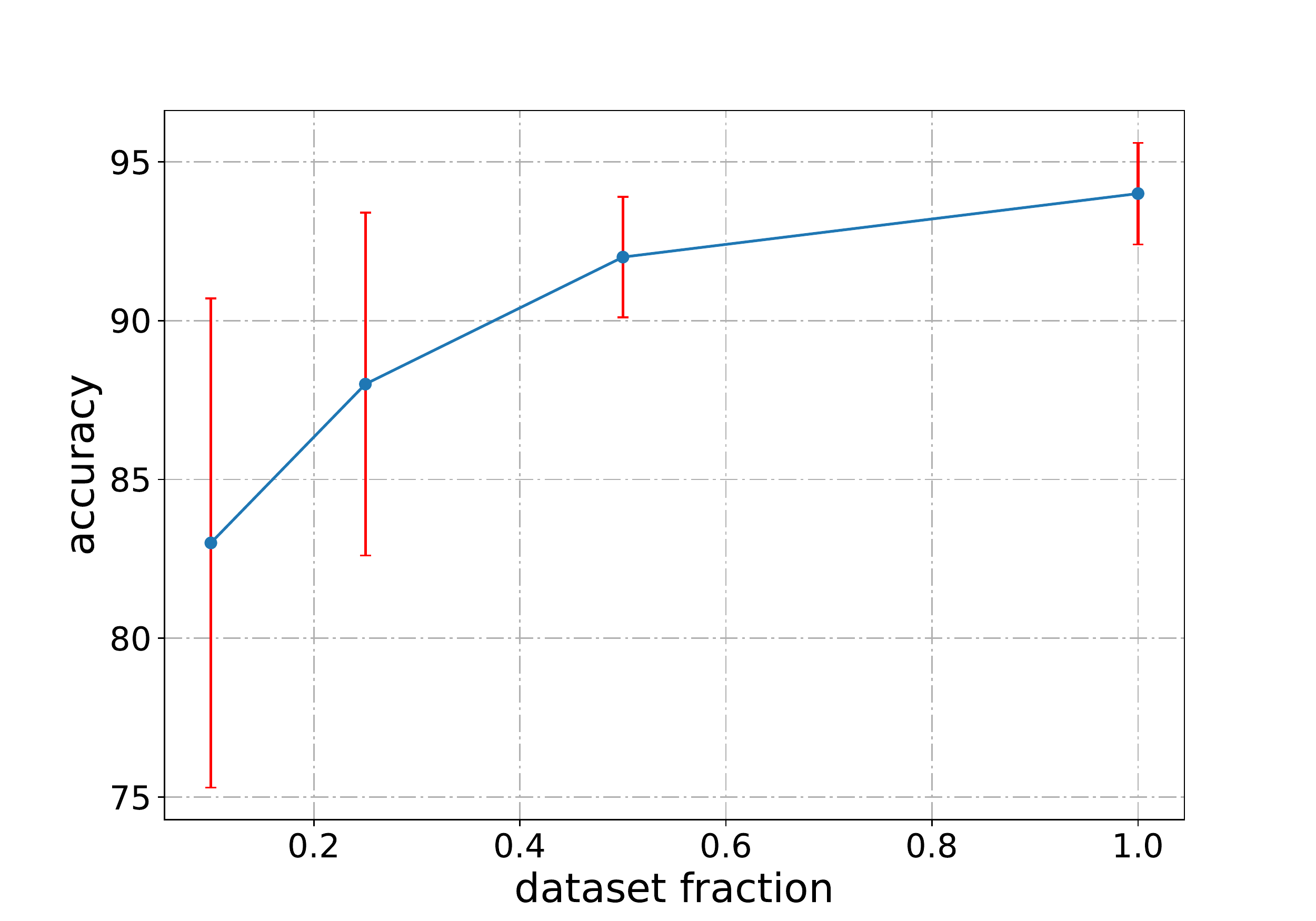}
  \caption{The LSTM RNN model accuracy when trained with fractions of (Dataset1 + Dataset2 + Dataset4). The standard deviation is indicated by the red bars.}\label{fig:acc_data_size}
\end{figure}

Figure~\ref{fig:acc_data_size} shows the model accuracies when trained with different dataset sizes. It shows that the model accuracy increases and becomes more stable (reduced variance) as it is trained with more data.

\subsubsection{Impact of model parameters on the accuracy}
\begin{figure*}
\centering
  \includegraphics[width=0.32\textwidth]{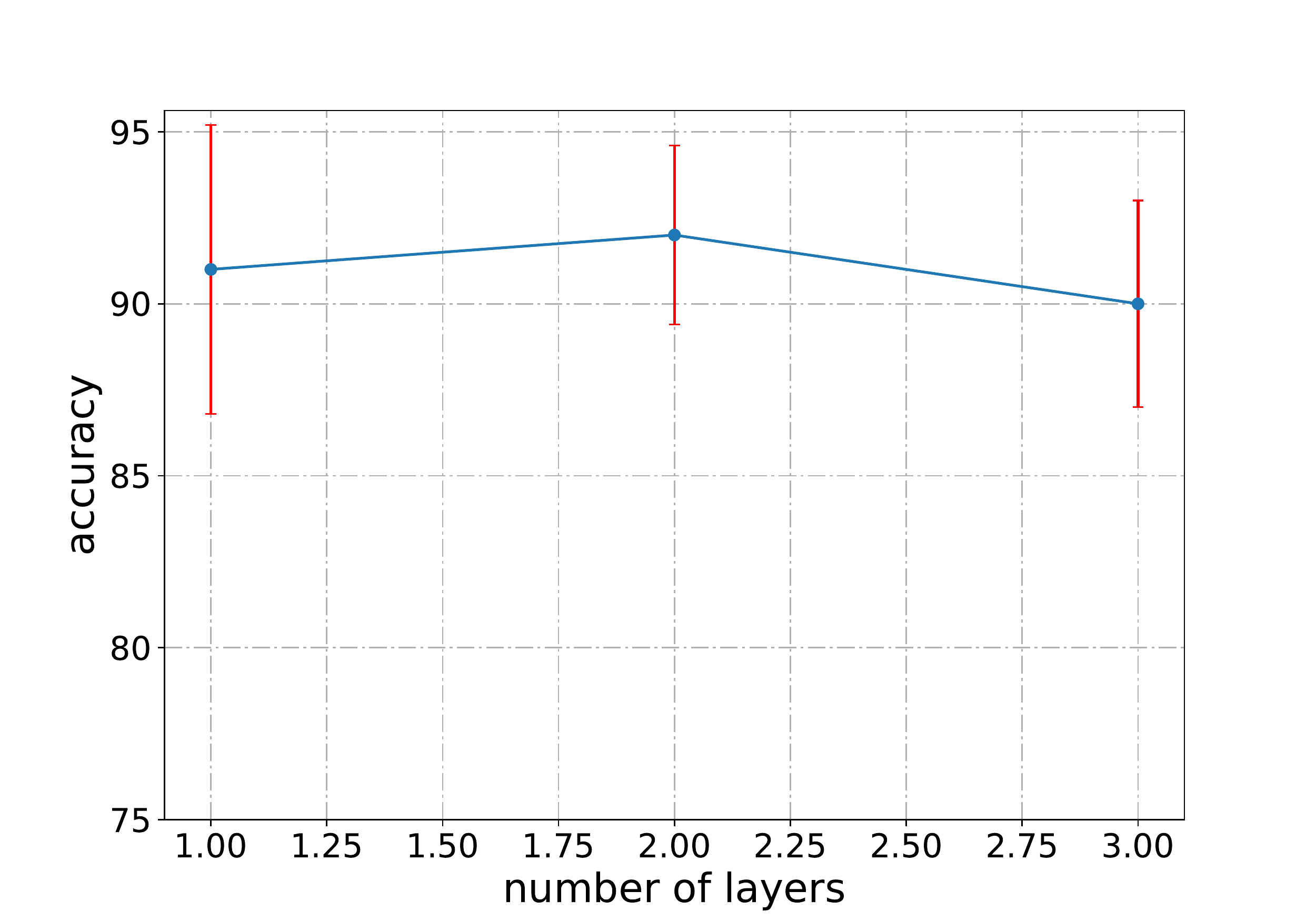}
  \includegraphics[width=0.32\textwidth]{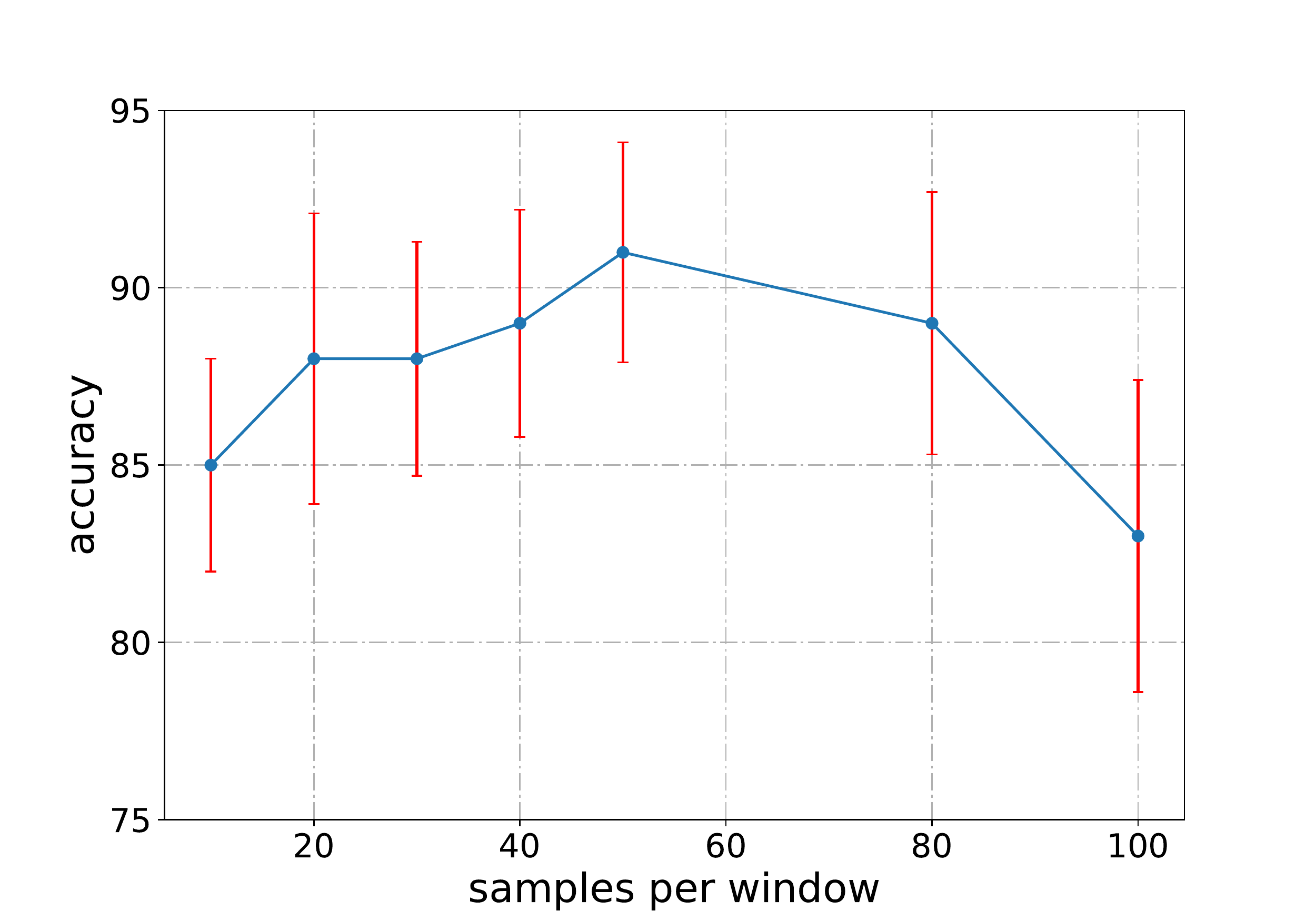}
  \includegraphics[width=0.32\textwidth]{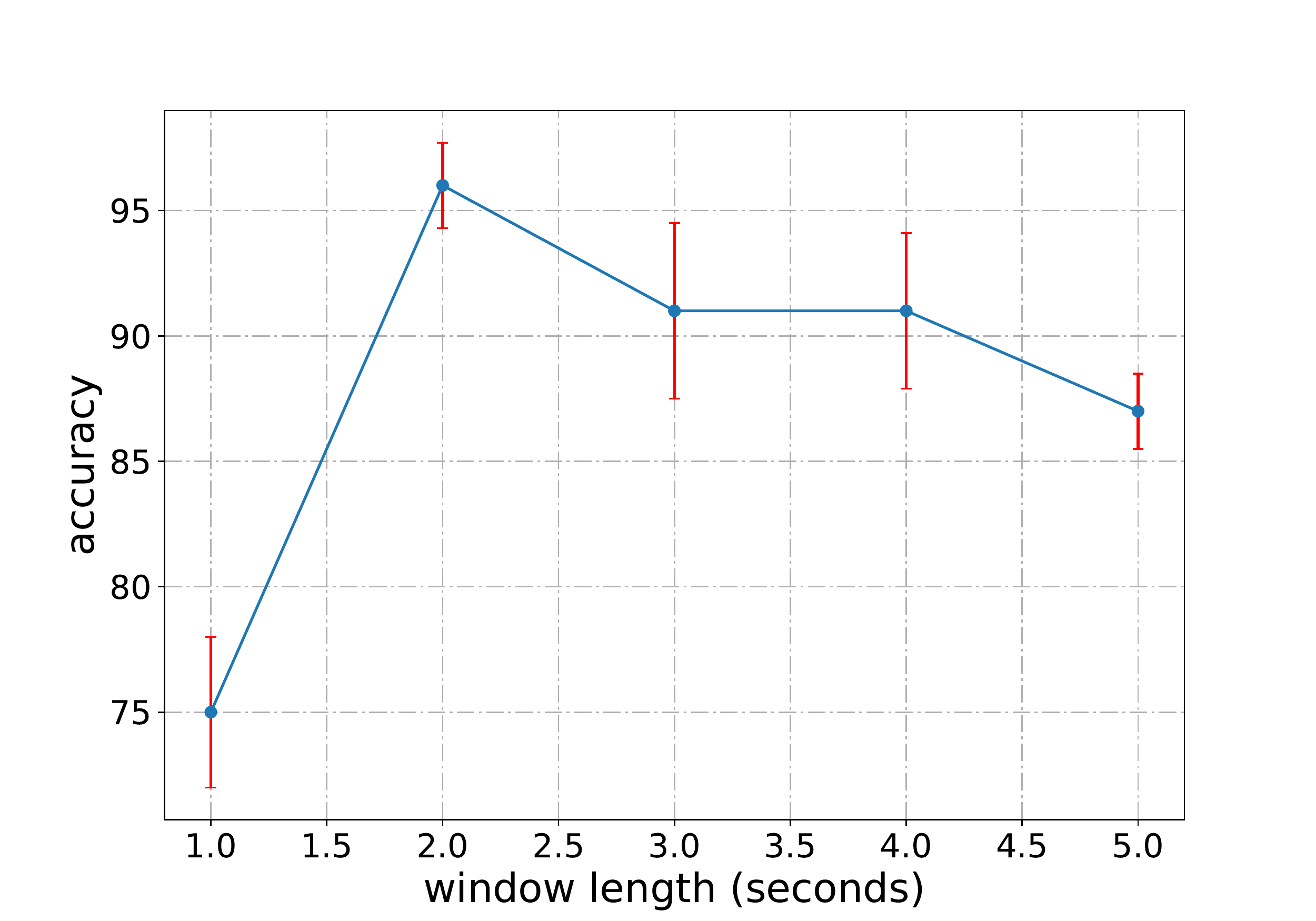}
  \caption{Prediction accuracy (in \%) as a function of the number of hidden layers in LSTM RNN (left), number of samples per prediction window (center), and the prediction window length (right). The standard deviation is indicated by the red bars.}
  \label{fig:acc_impact}
\end{figure*}

\paragraph{Model complexity}
As can be seen in Fig.~\ref{fig:acc_impact}, the experiments show that increasing the number of layers initially increases the model accuracy, but then decreases it, most likely due to overfitting.

\paragraph{Samples per window}
The results in Fig.~\ref{fig:acc_impact} shows that the model accuracy increases as the number of samples per window increases, but after a specific point the accuracy decreases. This decrease in accuracy can be explained by the increase in model complexity as the number of input samples increases (and hence increases the RNN time steps) which causes the model to overfit the training data.
\begin{figure}[t]
\centering
\vspace{-4mm}
\includegraphics[width=0.8\linewidth]{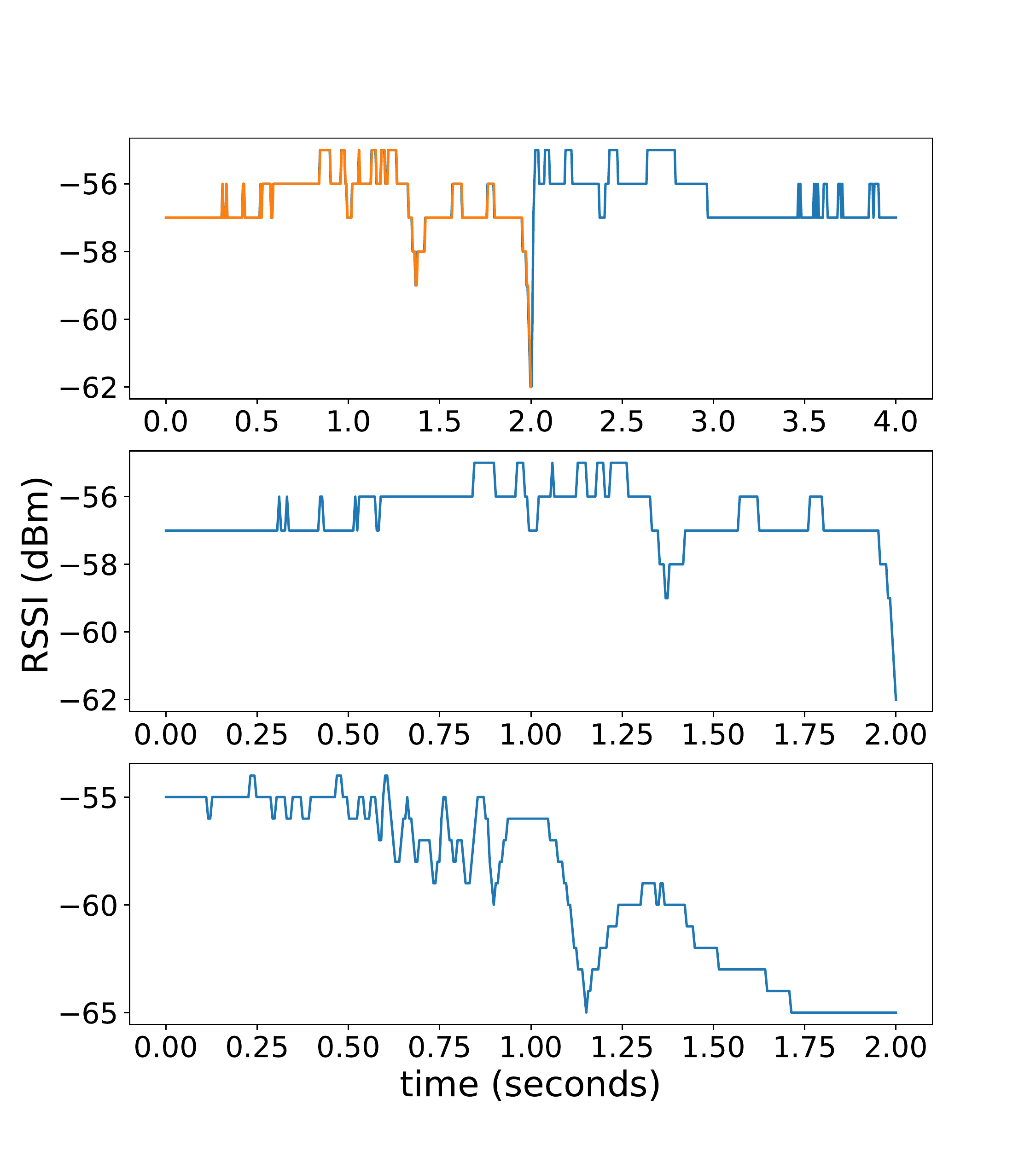}
  \caption{RSSI values recorded while a Swipe gesture is performed (top figure). The yellow highlighted part of the RSSI has been enlarged in the middle figure. The bottom figure shows a two seconds window of RSSI values recorded during a Push gesture, to illustrate the similarity between the Swipe and Push gestures in their RSSI fingerprints.}
\label{fig:short_context}
\end{figure}

\paragraph{Window size}
In Fig.~\ref{fig:acc_impact}, the impact of the prediction window length on the RNN model accuracy with the Dataset1 is shown. The model achieves the highest accuracy for the two seconds window length. However, online experiments showed that a two seconds window length provides a short context that results in a higher rate of confusion between gestures, see Fig.~\ref{fig:short_context}, and false positive predictions. Thus, a four seconds window is used in the online experiments.

\subsubsection{Impact of spacial setup on the accuracy}
\begin{figure}
\vspace{-2mm}
  \minipage{0.5\textwidth}
  \includegraphics[width=0.48\textwidth]{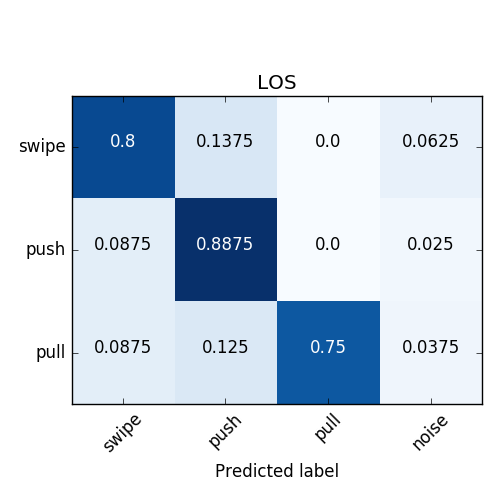}
  \includegraphics[width=0.48\textwidth]{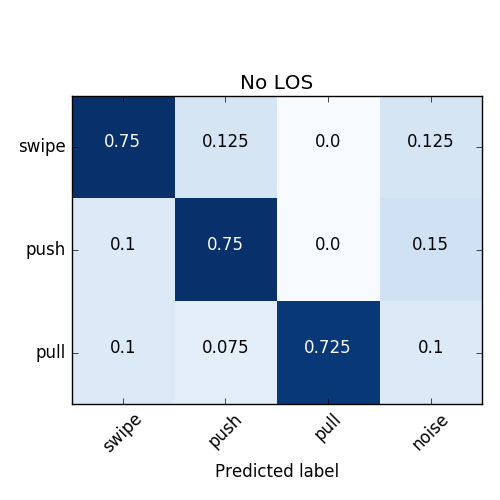}
  \endminipage\hfill
  \caption{Confusion matrix (accuracy) averaged over all LoS scenarios (left) and nLoS scenarios (right) for the online gesture classification method. 
 }\label{fig:conf_los}
\end{figure}

Table~\ref{table:online_experiments} and Fig.~\ref{fig:conf_los} summarize the various online experiments performed\footnote{Experiments made using the Wisture Android App. The RNN model had the same parameters as in Table~\ref{table:params}, except no dropout was used due to a limitation in the Tensorflow Android library.} and the obtained results\footnote{A video demonstration of sample results is available at \url{https://www.youtube.com/watch?v=rMv_bKkDtbU}.}. Accordingly it can be concluded that the solution: (1) generalizes to AP-smartphone spatial configurations that are different from the training ones, and (2) performs better in settings where there is a line-of-sight (LoS) between the smatphone and the AP ($81\%$ accuracy), than where there is no LoS ($74\%$ accuracy). The exception to this is the no-LoS online experiment number three, where an accuracy of $93\%$ was achieved, which is higher than all recorded LoS experiments accuracies. This might be due to the peculiarity of the Wi-Fi signal path between the smartphone and the AP in these cases. Refer to Sec.~\ref{sec:propagation} for more information on RF signal propagation.

Figure~\ref{fig:conf_los} shows that the system has approximately equal performance in recognizing the different gestures. Although Push has the highest average accuracy, the standard deviation (std) of the individual gestures show that, Push and Pull accuracies have a wider variation across the different test scenarios ($0.29$ std for Push and $0.3$ std for Pull) compared to the Swipe gesture ($0.15$ std). 
We believe the low accuracy achieved in experiment six is because of the fact that the AP and the smartphone were far apart (9 m), and the line-of-sight was blocked by a wall, a dishwasher and a fridge.
\begin{table*}
  \centering
  \begin{tabular}{l l l l l l}
    \hline
    No. & Spatial setup & Induction & Internet & Dataset & Accuracy\\
    \hline
    1 & Room A, LoS, 2m & $\surd$ & & Dataset1 & $85\%$ \\
    2 & Room A, LoS, 2m & $\surd$ & $\surd$ & Dataset2 & $67\%$\\
    3 & Phone room A, AP room B, no-LoS, 4.5m & $\surd$ & & Dataset4 & $93\%$\\
    4 & Room A, LoS, 2m & $\surd$ & $\surd$ & Dataset1 + Dataset2 + Dataset4 & $83\%$\\
    5 & Room B, LoS, 2.5m & $\surd$ & $\surd$ & Dataset1 + Dataset2 + Dataset4 & $90\%$\\
    6 & Phone room B, AP room C, no-LoS, 9m & $\surd$ & $\surd$ & Dataset1 + Dataset2 + Dataset4 & $30\%$ \\
    7 & Phone room A, AP room C, no-LoS, 8m & $\surd$ & $\surd$ & Dataset1 + Dataset2 + Dataset4 & $87\%$ \\
    8 & Phone room A, AP room C, no-LoS, 9.5m & $\surd$ & $\surd$ & Dataset1 + Dataset2 + Dataset4 & $87\%$ \\
    \hline
  \end{tabular}
  \caption{Online experiments summary. Each gesture was performed twenty times for experiments one to four, and ten times for the remaining experiments. The overall mean accuracy is $78\%$ (note, in experiment six, the Swipe gesture accuracy was $70\%$). Experiment eight was performed using a 2.4 GHz AP, while the data collection and all other experiments were performed using a 5 GHz AP.}
  \label{table:online_experiments}
\end{table*}

\subsubsection{False positive predictions}\label{sec:false_gestures}
To estimate the robustness of the system against RSSI changes caused by interfering background activities, the recognition application was left running for a period of thirty minutes on a table inside a room, while a person was in the same room typing on a computer (placed on the same table as the smartphone), and occasionally moving in the room. Table~\ref{table:false_positive} summarizes the results, which show an average false positive rate of $8\%$.
\begin{table}
  \centering
  \begin{tabular}{l l}
    \hline
    Gesture & Number of predictions (\%) \\
    \hline
    Noise (correct prediction) & 1652 ($92.1\%$)\\
    Swipe (False positive) & 61 ($3.4\%$)\\
    Push (False positive) & 62 ($3.5\%$)\\
    Swipe (False positive) & 18 ($1.0\%$) \\
    \hline
  \end{tabular}
  \caption{Distribution of predicted gestures when no hand gesture is performed over a period of thirty minutes.}
  \label{table:false_positive}
\end{table}

\subsubsection{Resource consumption}
The  Wisture app was found to utilize around $\sim$13\% of the total CPU time ($9\%$ system time + $4\%$ user time) when traffic induction is disabled. The utilization increased to $\sim$25\% ($13\%$ system time + $12\%$ user time) when induction is enabled. Note, the screen usage of the app are also counted into the resources (the screen was always on display to observe the experiment results). No CPU or power usage figures were reported by previous works.

\subsection{Summary}
Using the proposed recognition solution, the experiments showed that it is possible to detect and classify three different hand gestures with an accuracy of $78\%$, across a variety of spatial and traffic scenarios without modifying the smartphone hardware, operating system or firmware (on the contrary to the works in \cite{christoph2015}\footnote{In~\cite{christoph2015}, the wireless interface was in the "Monitor" mode, and hence captured all Wi-Fi traffic exchanged by all smartphones and APs in the smartphone vicinity. As a result, the available RSSI measurements increased and the solution achieved a high accuracy (~$90\%$). However, in addition to the need of a custom firmware, the "Monitor" mode comes at the cost of prohibiting all smartphone applications from sending or receiving traffic over Wi-Fi, and a likely increase in battery power consumption.},~\cite{sigg6813955},\cite{DBLP:journals/corr/NandakumarKG14}, and \cite{WiGest2015}). To the best of our knowledge, the work presented here is the first to demonstrate this. Note, although the works in~\cite{WiGest2015,DBLP:journals/corr/NandakumarKG14} demonstrated gesture recognition on unmodified devices, the solutions are tailored to a PC based implementation with higher resources than a smartphone\footnote{In \cite{DBLP:journals/corr/NandakumarKG14}, CSI data (currently available on Intel's Wi-Fi Link 5300 device driver) is used to identify hand gestures. In \cite{WiGest2015}, the RSS collection procedure is not described in detail.}.

\section{Limitations and Future work}
\subsection{Limitations}
\subsubsection*{Background activities}
The solution was sensitive to interference to background activities, but we did not analyze such impacts. They may affect the prediction performance in general, and false positives specifically. 

\subsubsection*{New gestures}
To support new gestures, a new dataset for the new gesture is needed. The developed recognition solution is trained with a dataset of $1000+$ sample gestures ($300+$ samples per gesture). The requirement of huge number of training samples can be seen as a limitation. However, we can see in Fig.~\ref{fig:acc_data_size} that the training data size can be reduced  to a tenth of the original set ($100+$ samples), with a modest decrease in accuracy to $83\%$ from $94\%$. 
\subsubsection*{Calibration to different Smartphones or devices}
Since the solution is based on training data from one particular smartphone, using the trained RNN on a different smartphone may result in a lower accuracy. To address this issue, further work is necessary to calibrate the training data to new smartphones or other devices (laptop, tablet, or a wearable device).

\subsection{Future work}\label{sec:imprv}
Leaving the recognition application running for a long time while no gesture is performed, increases the probability of false positive predictions and consumes the smartphone battery. This can be addressed by introducing a new mode of operation in which the recognition application waits for a special \textit{preamble gesture} that has two characteristics: (1) easy to separate from background noise; (2) requires small resource utilization to be detected. After a preamble detection mode, the application could detect further complex gestures. 

In our further works, we plan to use the preamble gesture concept, for instance with a sequence of Push and Pull gesture as a preamble. 
The results in Table~\ref{table:offline_ds_res} show that Push and Pull gestures can be recognized with an accuracy of $97\%$ even when no traffic is induced between the smartphone and the AP. Thus, during the preamble detection operation mode, traffic induction can be disabled, significantly reducing the power consumption and resource utilization of the application. Further, we will also investigate the influence of the distance and spatial characteristics between the smartphone and the hand making the gestures.

\section{Conclusion} 
In this work, we have demonstrated  that it is possible to recognize and classify contact-less moving hand gestures near smartphones without modification to the smartphone hardware or existing software. The proposed solution used a custom signal processing techniques, an artificial traffic induction approach, and a LSTM RNN based machine learning model to detect and classify the performed hand gesture from the smartphone Wi-Fi RSSI measurements. The solution achieved an average on-phone recognition accuracy of $78\%$ on average, and up to $94\%$ on specific datasets, when tested under several configurations (scenarios) including the ones that were different from the training data scenarios. We believe this accuracy qualifies the solution for non mission-critical applications.

In our future works, we aim to reduce the limitations of the solution such as vulnerability to interfering background activities (for instance by introducing a preamble detection mode), and calibrating the training data to new devices/gestures.

\section*{ACKNOWLEDGMENT}
We would like to thank Petter \"{O}gren for his invaluable input and advice on this work.

\bibliographystyle{IEEEtran}
\bibliography{references}	

\end{document}